\documentclass[aps,twocolumn,english]{revtex4}

\usepackage[T1]{fontenc}
\usepackage[utf8]{inputenc}
\usepackage{textcomp}
\usepackage{amsmath}
\usepackage{graphicx}
\usepackage{SIunits}
\usepackage{babel}
\usepackage{color}

\begin{document}
\title{Inducing micromechanical motion by optical excitation of a single quantum dot}

\author{J. Kettler$^{1,2}$, N. Vaish$^{1,2}$, L. Mercier de L\'epinay$^1$, B. Besga$^3$, P.L. de Assis$^{1,2,4}$, O. Bourgeois$^{1,2}$,
 A. Auff\`eves$^{1,2}$, M. Richard$^{1,2}$, J. Claudon$^5$, J.M. G\'erard$^5$, B. Pigeau$^1$,
O. Arcizet$^1$, P. Verlot$^6$, and J.P. Poizat$^{1,2}$
}

\affiliation{$^1$ 	Univ. Grenoble Alpes, CNRS, Grenoble INP, Institut NEEL,   France \\
$^2$ CNRS, Inst. NEEL,  "Nanophysique et semiconducteurs" group,  Grenoble,  France \\
$^3$	Univ Lyon, CNRS, Laboratoire de Physique, ENS de Lyon, France \\
$^4$		Gleb Wataghin Institute of Physics, University of Campinas, São Paulo, Brazil \\
$^5$ CEA, Univ. Grenoble Alpes, IRIG-PHELIQS, "Nanophysique et semiconducteurs" group, France \\
$^6$	School of Physics and Astronomy, University of Nottingham, United Kingdom }


\begin{abstract}
Hybrid quantum optomechanical systems offer an interface between a single two-level system and a macroscopical mechanical degree of freedom \cite{Rugar,Wilson-Rae,Lahaye,Rabl,O'Connell,Arcizet,Tian,Pirkkalainen,Treutlein,Teissier,Lee}.  
In this work, we build a hybrid system made of a vibrating microwire  
coupled to a single semiconductor 
quantum dot (QD)  via material 
strain. It was shown a few years ago,  that the QD excitonic transition energy can thus be modulated by the microwire motion \cite{Yeo,Montinaro}. 
We demonstrate here the reverse effect, whereby the wire is set in motion by the resonant drive of a single QD exciton with a laser modulated at the mechanical frequency. 
The resulting driving force is found to be almost 3 orders of magnitude larger than radiation pressure.
 From a  fundamental aspect, this state dependent force 
offers a convenient strategy to map the QD quantum state onto a mechanical degree of freedom. 

\end{abstract}

\maketitle

Optomechanics investigates the interactions between optical and mechanical degrees of freedom. The past two decades have witnessed the impressive success of cavity optomechanics \cite{Aspelmeyer}, whose principle relies on coupling the confined optical mode of an optical  cavity to a mechanical resonator. This approach enhances the effects of the optomechanical coupling via the  build up of the intra-cavity optical field, and led to landmark results, such as the observation of the zero point fluctuations of a mechanical resonator \cite{Chan}, and the preparation of entangled macroscopic mechanical states \cite{Riedinger}.

During this period, 
there have been theoretical proposals \cite{Wilson-Rae,Rabl} and experimental demonstrations \cite{Pirkkalainen,Treutlein}
 extending the paradigm of optomechanical cavities to so-called quantum hybrid systems, where the optical cavity is replaced by a single quantum optical emitter, typically modeled as a two-level system (TLS).
One of the major objectives for developing this  concept is the realization of a quantum interface between a qubit and a mechanical oscillator with important technological applications for quantum information and ultra-sensitive measurements \cite{Rugar,O'Connell,Satzinger,Chu}.
These devices also provide promising platforms for the emerging field of quantum thermodynamics \cite{Elouard}.
 
 Recently, hybrid systems operating in the microwave domain have witnessed impressive progresses \cite{Satzinger,Chu}. In optics however,  only a few experimental approaches have successfully implemented hybrid systems \cite{Arcizet,Tian,Yeo,Montinaro,Teissier,Lee}, where the coupling has been studied via the response of the TLS to motion, not the other way around.
  However, the reciprocal effect, corresponding to the backaction of a single quantum system on a macroscopic mechanical resonator, has remained elusive: Contrary to an optical  cavity \cite{Treutlein}, a TLS operates with no more than a single energy quantum at a time, which considerably reinforces the requirements on the magnitude of hybrid coupling.

\begin{figure}
\includegraphics[width=0.46\textwidth]{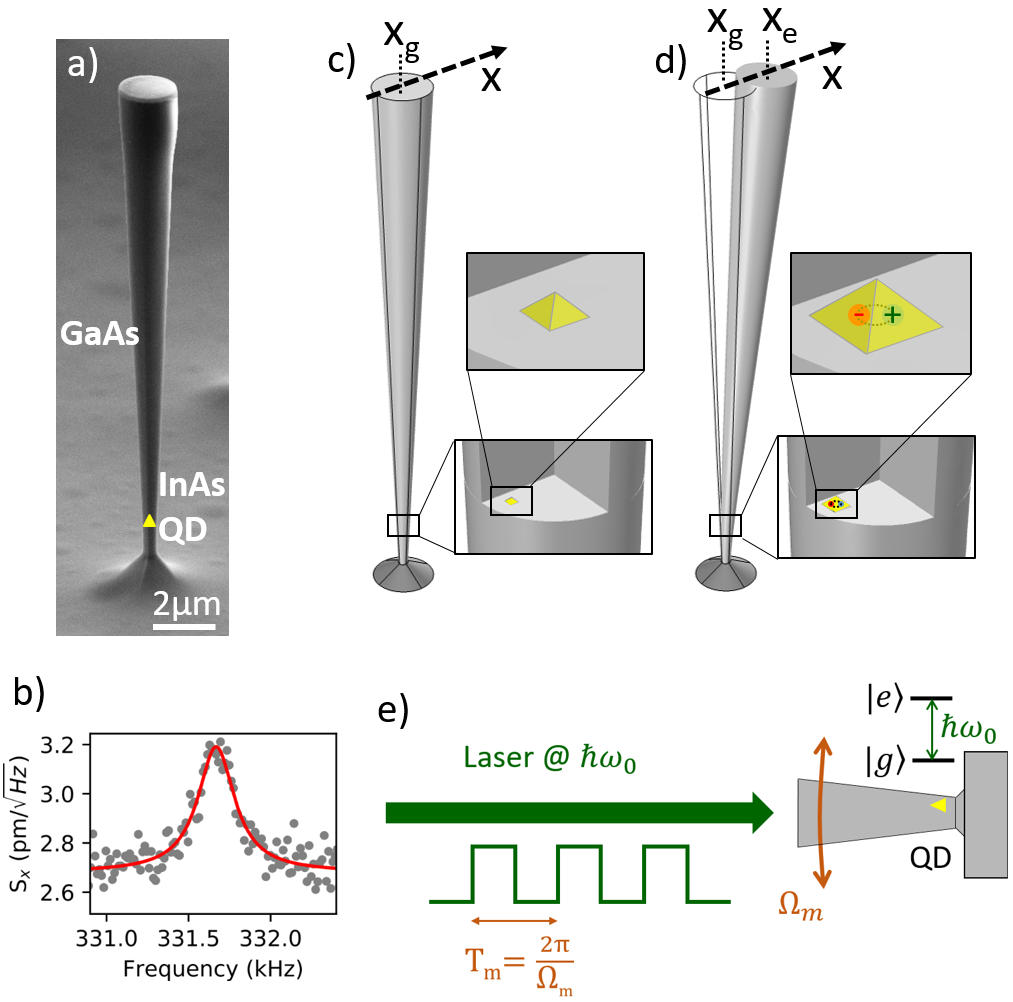}
\caption{\textbf{Principle}. \textbf{a}, Scanning electron microscope image of the GaAs photonic wire.  \textbf{b}, Brownian motion of the wire  at cryostat temperature of $T=5$ K (microwire temperature $T=25$ K, see SI) measured with a probe laser intensity  $P=100\;\mu$W as in the experiment.
\textbf{c}, When the QD is in its ground state, the wire rest position is x$_g$. \textbf{d}, When the QD is excited (hosts an exciton, represented as an orange electron and a green hole), its physical size is larger so that the rest position of the wire is displaced to x$_e$. 
\textbf{e}, Principle of the experiment. The QD is illuminated by a laser resonant with its optical transition at $\hbar \omega _0$. The laser intensity is modulated at the wire mechanical frequency $\Omega _m / 2 \pi$, and the wire motion is measured as a function of the laser detuning with respect to the QD transition.}
\label{Principle}
\end{figure}

In this work, we demonstrate  that a single optically driven quantum system can set in motion a solid-state mechanical oscillator of more than ten  micrometers long. As shown in Fig.~\ref{Principle}a, our hybrid system is composed of a single InAs quantum dot (QD) embedded close to the basis of a conical GaAs microwire \cite{Yeo,Munsch13}. The QD can be considered as a TLS with a ground state $\vert g\rangle$, and an excited state $\vert e\rangle$ corresponding to an additional trapped exciton. The energy of the TLS optical transition is coupled to the flexural vibration of the wire via strain-induced semiconductor band gap corrections \color{black} \cite{Wilson-Rae,Yeo,Montinaro,Metcalfe,Schulein,Golter,Carter18,Carter19}.
\color{black} Inspired by a recent proposal \cite{Auffeves}, we employ here optical excitation to modulate the exciton population at the mechanical frequency: the resulting periodic deformation of the QD then generates a flexural oscillation of the wire.

The dynamics of this hybrid strain-coupled system is  described by the  parametric Hamiltonian,
\begin{equation}
H=\hbar\omega_0 \frac{(\hat{\sigma}_z+1)}{2} + \hbar \Omega_m (\hat{b}^\dagger \hat{b}+1/2) + \hbar g_m  \frac{(\hat{\sigma}_z+1)}{2}(\hat{b}+\hat{b}^\dagger),
\label{Hamiltonian}
\end{equation}
where  $\hbar\omega_0$ is the
QD transition energy for the wire at rest ($\hbar\omega_0 \simeq 1.35$ eV, wavelength $920$ nm), and
$\hat{\sigma} _z=\vert e \rangle\langle e \vert - \vert g \rangle\langle g \vert$ is the Pauli operator describing the population of the QD states.
$\Omega_m/2\pi$ is the mechanical eigenfrequency of one of the wire fundamental flexural modes ($\Omega_m/2\pi = 330$ kHz, see 
Fig.~\ref{Principle}b), and $\hat{b}$ the phonon annihilation operator  of this mode. 
The last term of Eq.(\ref{Hamiltonian}) describes the strain-mediated coupling, and can be rewritten as $\hbar g_m  \frac{(\hat{\sigma}_z+1)}{2}\frac{\hat{x}}{x_{\mbox{\scriptsize{zpf}}}} $. Here, $\hat{x}=x_{\mbox{\scriptsize{zpf}}}(\hat{b}+\hat{b}^\dagger) $ is the top facet position of the wire, with $x_{\mbox{\scriptsize{zpf}}} =\sqrt{\hbar/2m\Omega_m} $  the corresponding zero point fluctuations and $m$  the microwire  effective mass associated with the flexural mode. The coupling strength reads $g_m = \frac{ d \omega_0 }{d x} x_\text{zpf}$. Following \cite{Yeo}, we have measured $\frac{d \hbar \omega_0}{d x} = 10 \;\mu$eV/nm  from the QD photoluminescence lineshape broadening when flexural motion is excited (see SI), while $m = 32 $ pg and $x_{\mbox{\scriptsize{zpf}}} = 28 \: $fm  are deduced from Brownian motion measurements  (see Fig.~\ref{Hamiltonian}b and SI). For the chosen QD, this yields $g_m / 2 \pi = 68\: $kHz.


The interaction between the microwire and the QD relies on the strain difference that the material surrounding the QD experiences when the QD hosts a single exciton \cite{Besombes}. 
It follows that the rest position x$_e$ of the mechanical oscillator for an excited QD   is different from the rest position x$_g$ corresponding to an empty QD (see Fig.~\ref{Principle}c,d and SI). The difference between these two positions is given by
\begin{equation}
\mbox{x}_e-\mbox{x}_g = -2\frac{g_m}{\Omega_m}   x_{\mbox{\scriptsize{zpf}}}.
\end{equation}
A consequence of this situation is that, when the QD is in the ground state and is optically brought in the excited state on a time scale much faster that $2\pi / \Omega_m$, the  QD induces on the wire a static  force 
\begin{equation}
F_{\mathrm{QD}}=k(\mbox{x}_e-\mbox{x}_g),
\end{equation}
 with $k=m\Omega_m^2$ the oscillator stiffness.

In our experiment, the  QD is driven by a laser whose intensity is chopped with a $50\%$ duty cycle at the wire's mechanical frequency  
$\Omega_m/2\pi$. 
The QD exciton decay time typically amounts to
 $\tau_{\mathrm{QD}}=1$ ns   \cite{Munsch17},  about 3 orders of magnitude faster  than the mechanical oscillator period ($\sim 0.3\;\mu$s).
 Thus, the force generating the mechanical motion only depends on  
 the average  QD  population measured by $\left\langle\hat{\sigma}_z\right\rangle$ over half a period. When the laser is "on" (resp "off") the maximum average excited state population is $n_e=1/2$ (resp $n_e=0$), so that  $\left\langle \hat{\sigma}_z\right\rangle =0$
 (resp. $\left\langle \hat{\sigma}_z\right\rangle =-1$). 
So the wire rest position is periodically displaced from x$_g$ to $\mbox{x}_g +(\mbox{x}_e-\mbox{x}_g)/2$, leading to the excitation of a QD induced wire motion (fig.~\ref{Principle}e) with a root mean square (rms) amplitude at resonance (see SI):
\begin{equation}
x_{\mathrm{QD}}=\frac{\sqrt{2}}{\pi} \eta \frac{g_m}{\Omega_m} Q x_{\mbox{\scriptsize{zpf}}},
\label{xQD}
\end{equation}
where $\eta$ is an efficiency factor accounting for the fact that photons reaching the QD may not all excite 
it due e.g. to spectral wandering of the transition line,
 and $Q$ is the wire mechanical quality factor ($Q= 1650$). For $\eta=1$, our system parameters lead to $x_{\mathrm{QD}}^{\eta=1}=4.4$ pm,  whereas the  rms spread of the Brownian motion at a temperature  $T=25$ K is $x_{Br}=x_{\mbox{\scriptsize{zpf}}}\sqrt{2 k_B T/\hbar \Omega_m}=50$ pm.

\begin{figure}
\includegraphics[width=0.47\textwidth]{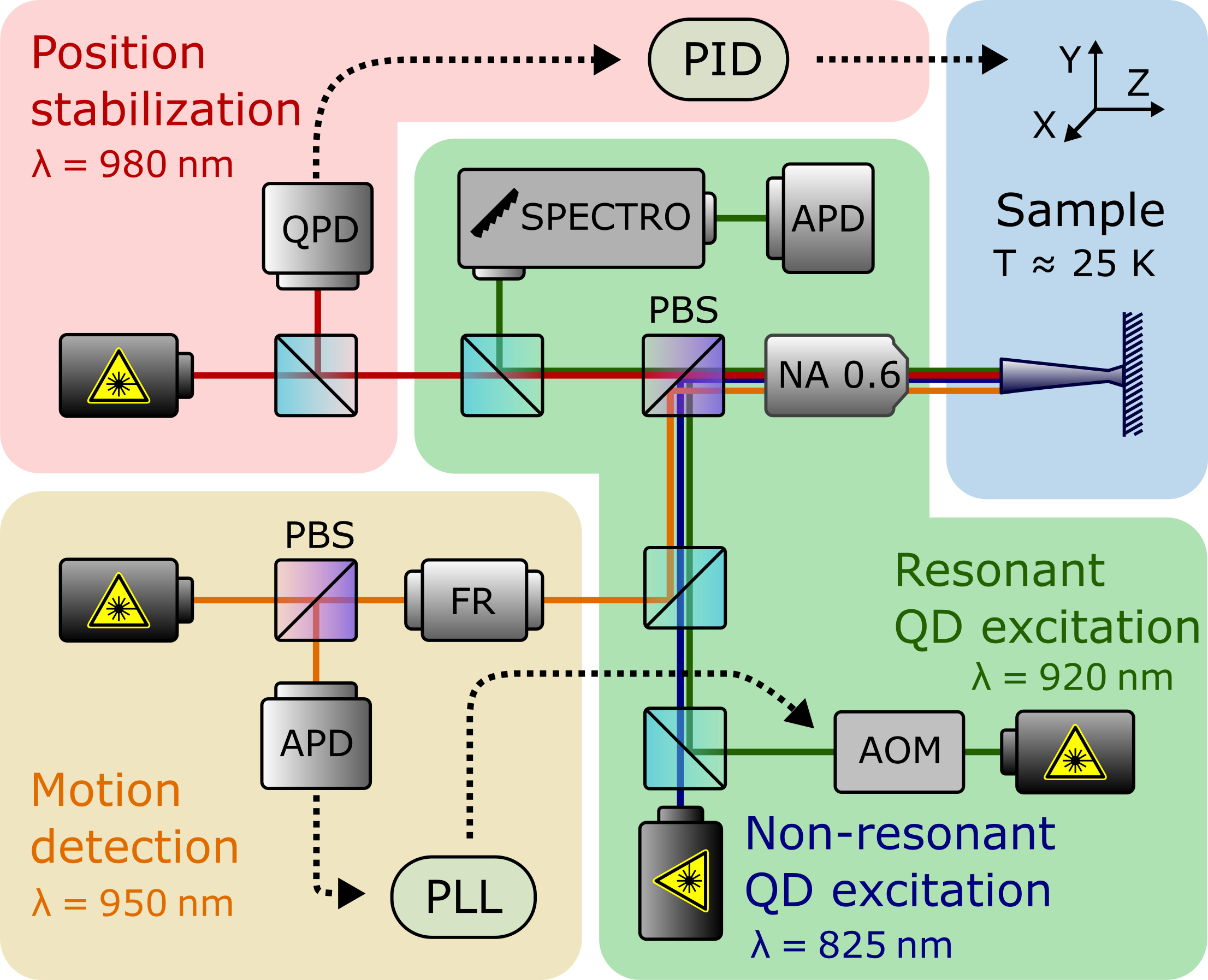}
\caption{\textbf{Experimental set-up}. The sample is anchored on the cold finger of a helium flow cryostat. The sample temperature is $T=25$ K. Resonant QD excitation (green background) is performed by an external cavity diode laser at $\lambda=920$ nm, which is intensity modulated at the wire resonant mechanical frequency by an acousto optical modulator (AOM). To remain on resonance, the modulation frequency is locked  on the photothermally actuated motion via a phase lock loop (PLL). A non resonant diode laser at $\lambda=825$ nm is used for preliminary photoluminescence characterization, and as a "cleaning" laser for the resonant excitation (see Methods). 
Fluorescence photons are counted by an Avalanche  Photodiode at the output of a high resolution ($12\;\mu$eV) spectrometer. Motion detection (orange background) is carried out with an external cavity diode laser at $\lambda=950$ nm. Its position dependent reflexion from the  wire top facet is extracted using a Faraday rotator and sent to a photodiode whose signal is demodulated with a lock-in amplifier (see Methods). The position stabilization (pink background) is performed with a diode laser at $\lambda = 980$ nm, whose reflexion from the wire top facet is detected on a quadrant photodiode feeding a Proportional Integral Derivative (PID) servo controlling  the microscope objective x-y position.
Lasers at $\lambda=920, \; 950, \;980$ nm are intensity stabilized. 
}
\label{Exp_Setup}
\end{figure}

The experimental set-up (see Fig.~\ref{Exp_Setup}) combines a resonant fluorescence set-up with a motion measurement scheme (see Methods).  
The excitation laser is intensity modulated by an acousto-optical modulator (AOM) at the mechanical frequency with an on-off square function. Its continuous wave intensity before the microscope objective is  $150$ nW, slightly above the QD saturation power of $100$ nW. 
 A characteristic property of QD induced motion
is that it features an amplitude  proportional to the QD excited state population, which is monitored by the QD fluorescence. To reveal this signature, we scan the excitation laser frequency across the QD resonance while measuring simultaneously the QD fluorescence and the wire motion (see Methods). The latter  is demodulated at the driving mechanical frequency to determine its two quadratures. 
The incoherent Brownian motion  is averaged out by acquiring data for more than $20$ hours, while laser intensities and microwire position are actively stabilized (see SI).

\begin{figure*}
\includegraphics[width=0.8\textwidth]{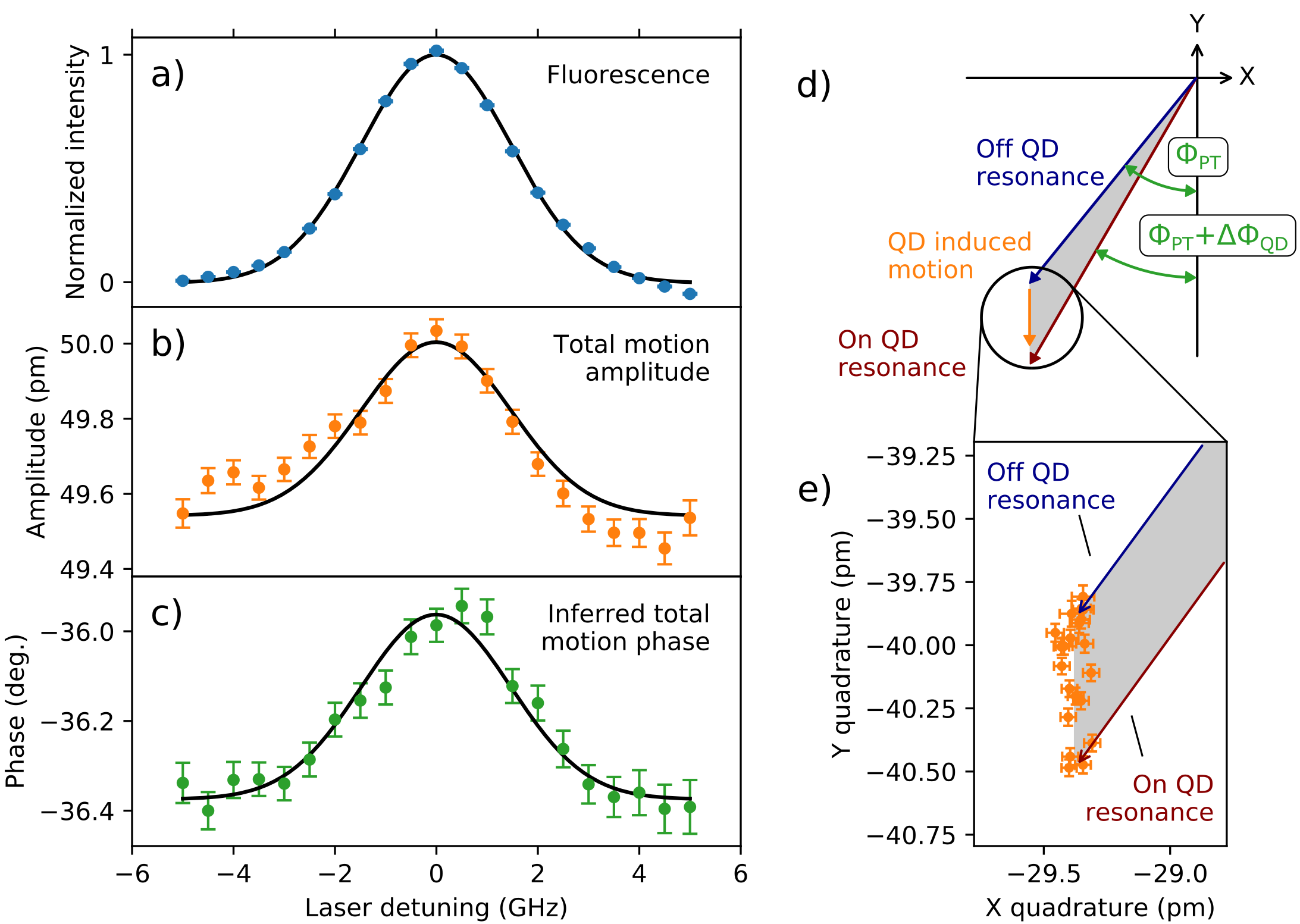}
\caption{\textbf{Experimental results}. \textbf{a}, QD  fluorescence, \textbf{b}, total motion rms amplitude, and \textbf{c}, total motion phase shift, are displayed as a function of the resonant laser detuning with respect to the QD transition. The fluorescence data is fitted with a Gaussian whose centering and width are kept for the fitting of the motion amplitude and the phase shift. In \textbf{c}, the phase shift is inferred from the frequency shift the PLL has to apply to maintain the set phase  (see Methods and SI). 
\textbf{d}, \color{black} This sketch represents the coherent component of the wire motion  in the quadratures plane (rotating frame). $X$ and $Y$ are the rms values of the coherent  motion quadratures, 
defined as $x_c(t)=\sqrt{2}( X \cos \Omega_m t -  \sin \Omega _m t )$. The phase reference is the laser modulation $I_{las}(t)= I_0 \cos \Omega_m t$. A phase delay is thus described with  a negative angle. \color{black} 
 In blue is shown the total motion vector  when the laser is off QD resonance. This corresponds to the photothermal (PT) motion only. The orange vector represents the QD induced motion at QD resonance.  The red vector displays the total motion at  QD resonance. 
It is recalled that for a drive with instantaneous action, the motion at the mechanical resonance exhibits a $-90^\circ$ phase shift. The PT motion features  an extra $\Phi_{PT}=-36^\circ$ phase shift. 
In \textbf{e} are displayed  experimental data corresponding to a zoom on the effect of the QD induced motion on the total motion. The photothermal background motion is represented by the blue vector.  Each data point represents the total motion (ie photothermal + QD induced)  corresponding to a different laser detuning across the QD resonance as in \textbf{a}-\textbf{c}.  
For all the graphs of this figure, error bars are the standard error of the mean.
}
\label{Exp_results}
\end{figure*}

Experimental results are shown in Fig.~\ref{Exp_results}. 
The QD fluorescence is plotted as a function of laser detuning in Fig.~\ref{Exp_results}a, together  with  the wire motion amplitude in Fig.~\ref{Exp_results}b.
The averaged motion amplitude features a peak which perfectly matches the  resonance condition between the laser and the QD transition. This constitutes a first signature of the QD induced motion. The motion peak   lays on top of a 50 pm non-resonant background. Since the laser power is sufficiently low to avoid any significant radiation pressure or gradient force, we attribute this background to  photothermal actuation of the wire [24] (see Methods).

A second signature of the QD induced motion is related to its phase. As seen in Fig.~\ref{Exp_results}c,e the photothermal motion features a  phase delay $\Phi_{PT}=-36^\circ$ between the excitation and the force.
 This delay is attributed to heat conduction and dissipation within the wire (see SI). On the other hand, the QD induced force is expected to be quasi-instantaneous as it follows the comparatively fast QD dynamics ($\tau_{\mathrm{QD}} \ll 2\pi/\Omega_m$).  The total coherent motion resulting from the coherent addition of these two actuation mechanisms experiences  therefore a phase shift $\Delta \Phi_{QD}$ with respect to the photothermal motion alone (see Fig.~\ref{Exp_results}d).   This phase shift is indirectly measured via a phase locked loop (PLL)  used to track the mechanical frequency (see Methods and SI for details). As seen  in  Fig.~\ref{Exp_results}c the inferred averaged motion phase shift exhibits a peak at QD resonance, whose origin is the QD induced motion. 
The rms coherent motion vector extremities  for  detunings scanned across the QD resonance are displayed in  Fig.~\ref{Exp_results}e   in the quadratures plane.  The vertical orientation of these points indicates that the QD induced force follows instantaneously the laser modulation,  in line with the fact that the  QD induced motion is caused by a strain effect governed by the fast QD dynamics.

The experimental rms amplitude of the QD induced motion is $x_{\mathrm{QD}}=0.6 \pm 0.1$ pm as seen in Fig.~\ref{Exp_results}e.
This result matches the expected value, 
\color{black} by adjusting $\eta$ to $\eta=0.15$ in Eq.(\ref{xQD}). \color{black}
This non-ideal $\eta$ value is attributed to the fast (faster than $1$ s) spectral diffusion of the resonantly excited QD transition leading to an effective blinking with $15\%$ of "on" periods as already observed with similar QDs \cite{Munsch17}. 

 The above measured displacement amplitude corresponds to a force $F_{\mathrm{QD}}^{\mathrm{exp}}=\frac{m\Omega_m^2}{Q}\times x_{\mathrm{QD}}\simeq 50\,\mathrm{aN}$. This is almost 3 orders of magnitude larger than the radiation pressure force, which corresponds to the momentum exchange between photons and the nanowire over the exciton lifetime, $F_{\mathrm{rad}}\simeq\frac{2\hbar k_l}{\tau_{\mathrm{QD}}}\simeq 0.1\,\mathrm{aN}$, with $\tau_{\mathrm{QD}}\simeq 1\,\mathrm{ns}$ the exciton lifetime and $k_l$ the light wave-vector (see Methods).

Remarkably, this strong single-photon force enhancement compared to radiation force is granted while the underlying photon-exciton conversion process is preserving the photon number. This essential property makes our approach intrinsically suitable to quantum photonics engineering, including coherent processing of nonclassical quantum superpositions of QD states, and to the use of mechanics as a quantum bus for storing and routing entangled QD states. 
\color{black}Such compelling applications require important technological improvements. Simulations show that optimizing and miniaturizing the geometry enables to increase both mechanical resonance frequency and hybrid coupling rate \cite{Artioli}. This strategy yields a triple benefit: i) At higher frequencies, the photothermal effect becomes much slower than the mechanical period and therefore vanishes. ii) Concurrently, this allows the system to operate in the resolved sideband regime \cite{Garcia-Sanchez,Eichenfield}, which enables the preparation and manipulation of quantum motion states \cite{Wilson-Rae}. iii) Furthermore, higher coupling rates open the perspective to reach the strong coupling regime, were the coupling rate exceeds all decoherence rates.

Note that, our result can be readily extended to existing systems, starting with waveguide embedded single photon sources  \cite{Lodahl}, which may host ultra-high frequency mechanical degrees of freedom \cite{Gavartin,Esmann}. Other potential high-frequency candidates include hybrid phononic nanostrings \cite{Vogele}, surface acoustic waves coupled to quantum dots and NV centers \cite{Metcalfe,Schulein}, and hybrid photonic crystal cavities \cite{Sun}. 

\color{black}

These improvements notwithstanding, our experiment can already be interpreted and exploited from the more fundamental standpoint of quantum thermodynamics, the whole process being reminiscent of a nanoengine. The quantum emitter converts the energy incoherently provided by the light field (heat) into coherent mechanical motion (work). Importantly, the bath is out-of-equilibrium, which opens novel scenarios for quantum and information thermodynamics \cite{Rossnagel,Klaers}.

To summarize, we have been able to set in motion a micro-oscillator of mass $0.1$ ng by optically manipulating the quantum state of a single quantum object. The motion is generated by periodically driving the excitonic population of an embedded quantum dot, which is coupled via strain to the position of the oscillating wire. This built-in coupling is extremely robust to any external perturbation and features a high level of integration, making the next generation of hybrid photonic nanowires a potential solid-state building block for quantum photonics circuitry.

\section*{Acknowledgements}

JK, NV, AA, JC, JMG, PV and JPP are supported by the French National Research Agency (project QDOT ANR-16-CE09-0010). NV is supported by Fondation Nanosciences.
PLdA thanks Université Grenoble Alpes and CNRS for supporting visits as invited scientist.  
AA is supported by the Agence Nationale de la Recherche under the Research Collaborative Project Qu-DICE (ANR-PRC-CES47).
MR  is supported by the Agence Nationale de la Recherche under the Research Collaborative Project QFL (ANR-16-CE30-0021).
BP is supported by the Agence Nationale de la Recherche under the project QCForce (ANR-JCJC-2016-CE09). OA acknowledges support from ERC Atto-Zepto. PV acknowledges support from the ERC StG 758794 "Q-ROOT". Sample fabrication was carried out in the “Plateforme Technologique Amont” and in CEA/LETI/DOPT clean rooms.

\section*{Author contributions}

JK and NV performed the experiments with the help of BB and PLA. JK and LML wrote the experimental codes. JK performed the data analysis. OB did the photothermal analysis of the system. AA and MR provided theoretical support. JC and JMG designed and fabricated the samples. JK, BP, OA, PV and JPP proposed the experimental procedures. JPP supervised the project and  wrote the manuscript with the help of AA, MR, JC, JMG, BP, OA and PV.

\section*{Competing interests}

The authors declare no competing interests.

\section*{Methods}

\subsection*{QD fluorescence detection}

The QD fluorescence (wavelength around $920$ nm) is detected by a photon counting avalanche  photodiode at the output of a $1.5$ m focal length spectrometer equipped with a $1200$ grooves/mm grating (resolution $12\;\mu$eV). The resonant laser reflexion is attenuated by a cross polarizer scheme. To  further reject the resonant laser light,  the spectrometer is adjusted on the high energy phonon side band $0.5$ meV away from the QD line \cite{Besombes}. Note that these high energy phonons ($0.5$ meV $=120$ GHz) are not at all related to the wire motion but to bulk semiconductor crystal thermal excitations.  The phonon side band signal is lower than the zero phonon line, but the laser background is almost completely removed leading to a greatly improved signal to noise ratio (see SI). 

Additionally, as already observed by other groups   \cite{Majumdar}, the sample must be illuminated by a weak power ($100$ nW) non resonant (wavelength $825$ nm) laser to saturate defects around the QD to reduce spectral diffusion and  observe resonance fluorescence in good conditions. 

Integrated over several hours, the QD linewidth is in the $100\; \mu$eV range owing to slow spectral wandering on a time scale of several minutes. After proper data processing (see SI), we can  reduce the effects of slow spectral wandering  to reach a linewidth of $12 \;\mu$eV (see Fig.~\ref{Exp_results}a).

Note that polarization sensitive photoluminescence spectroscopy of the chosen QD has shown that the line of interest comes from a charged exciton (trion).

\subsection*{Wire motion detection}

The wire motion is detected via the reflexion of a laser focused on the edge of the wire top facet, so that the  wire motion  modulates the  reflected intensity. This probe laser is a shot-noise limited  external cavity diode laser.
Its wavelength ($950$ mn) is chosen so that it can be efficiently filtered from the QD fluorescence light using tunable edgepass  interference filters, and  its energy is below the QD transition energy to minimize its impact on  QD excitation. The mechanical signal to noise ratio (in power) scales linearly with the probe laser intensity. However  the detection of the QD resonance fluorescence is degraded for probe beam intensities exceeding $100\;\mu$W. We are thus limited to this intensity of $100\;\mu$W.
This low light level requires the use of a high gain avalanche photodiode (APD) that enables us to measure the Brownian motion at a cryostat temperature of $T=5$ K  (sample temperature of $T=25$ K, see SI) in good conditions (see fig.~\ref{Principle}b). 
  Finally, for the displacement measurement to be quantitative, we carefully calibrate  the  APD voltage change $dV$ produced by a known static displacement $dx$ of the probe laser with respect to the wire (see SI).

\subsection*{Photothermal motion}

The resonant laser is at an energy of $E_l=1.35$ eV (wavelength $920$ nm) well below the GaAs gap at $T=25$ K ($E_g=1.518$ eV, wavelength:  $\lambda_g=817$ nm).
It is nevertheless very weakly absorbed along the  wire, mainly by surface defects. 
Its on-off modulation at mechanical frequency leads to  periodic heating   and thus  deformation of the non-perfectly symmetrical wire. 
From the QD line shift when the laser is "on", the  temperature increase caused by light absorption has been estimated  to be less than $0.01$K. 
This QD independent driven motion has a root mean square amplitude  of $x_{PT}=50$ pm  and features a phase delay $\Phi_{PT}=-36 \pm5 ^\circ $ with respect to a  motion that would respond instantaneously to the excitation laser modulation. This phase delay is well accounted for by estimating the thermal time response of the wire (see SI).

Thanks to this photothermal actuation we can lock the laser modulation frequency to the wire mechanical resonance using a Phase Locked Loop (PLL) to cope with the slow mechanical frequency drift attributed to wire icing.  Additionally,  the PLL in-loop frequency signal is used to infer the total motion phase shift (see below and SI).

\section*{Experimental procedure}

The bottom of the wire contains a layer of about $100$ self-assembled QDs  (cf Fig.~\ref{Principle}a and SI). Different QDs exhibit different transition energies, so that resonant excitation allows us to address a single QD. The asymmetrical section of the wire gives rise to two non-degenerate fundamental flexural mechanical modes. The  chosen QD  is  off-centered (about one fifth of the radius) so that it experiences strain as the wire oscillates along  one of the two linearly polarized fundamental flexural modes \cite{deAssis}, but not too far out to remain well coupled to the guided optical mode of the wire, and immune to surface effects.

The experimental procedure consists in scanning the intensity modulated resonant laser wavelength across the QD transition, while recording the motion of the wire.  
The QD induced motion being smaller than the Brownian motion ($x_{\mathrm{QD}}/ x_{Br}\sim 1/100  $), the signal must be integrated over time to average out the Brownian motion (see SI). The chosen strategy is to carry out more than $1000$ scans of one minute over a duration of about 20 hours. The scan data are postprocessed before the final averaging allowing to reduce the effects of technical noises and QD instabilities (see SI).

 To ensure stability of the experiment during this time, the power of all lasers is stabilized, the thermal drift of the wire position with respect to all lasers is tracked using an extra laser (wavelength $980$ nm) reflected on the wire  towards a quadrant photodiode (see SI), and the mechanical frequency is tracked with a phase locked loop (PLL), (see SI).

\subsection*{Phase shift measurement}

During the experiment the  mechanical drive frequency is locked to the wire resonance using a Phase Locked Loop (PLL). The total motion phase change $\Delta \Phi_{QD}$ caused  by the QD induced motion is compensated by  the PLL via the shift of its driving frequency by $\Delta \Omega /2\pi$ in order to remain on the set phase. For frequency change smaller than the mechanical linewidth (i.e. $\Delta \Omega \ll \Omega_m/2Q$)   the phase to frequency conversion factor  at mechanical resonance is $d\Phi / d \Omega =2Q/\Omega_m$, so that  $\Delta \Phi_{QD}  =(2Q/\Omega_m) \Delta \Omega $. 
This allows us to infer the QD induced motion phase change $\Delta \Phi _{QD}$ from the measured 
 frequency change $\Delta \Omega /2 \pi$  (see SI for more details), as plotted as the green dots in Fig.~\ref{Exp_results}c.

\subsection*{Radiation pressure force}

In the main text, we compare the exciton induced force $F_{\mathrm{QD}}$ to the radiation pressure generated by a laser beam perpendicular to the wire axis perfectly reflected at the very tip of the wire. This force is given by
 $F_{\mbox{\scriptsize{rad}}}=\frac{2\hbar k_l}{\tau_{\mathrm{QD}}}$,
with $k_l$ the laser wave-vector, and $\tau_{\mathrm{QD}}=1$ ns the QD lifetime. 

\section*{Data availability}

 Data  are available from the public repository  \href{url}{https://zenodo.org/record/4118790#.X5H6jZrgqV4}.

\section*{Code availability}

Data processing code  is available from the public repository  \href{url}{https://zenodo.org/record/4118790#.X5H6jZrgqV4}.

\clearpage

{\LARGE \textbf{Supplementary information}}

\section{Magnitude of the QD induced motion}
\subsection{Expression of the force}
The system  is made of a conical GaAs photonic wire hosting an optically active InAs quantum dot (QD) behaving as a two-level system (TLS) with ground state $\vert g\rangle$, and excited state $\vert e \rangle$ separated by an energy $\hbar\omega_0$. The  QD  is not positioned at the center so that it experiences strain as the wire oscillates at frequency $\Omega_m/2\pi$ resulting in the following  parametric Hamiltonian:
\begin{equation}
H=\hbar\omega_0 \frac{(\hat{\sigma}_z+1)}{2} + \hbar \Omega_m (\hat{b}^\dagger \hat{b}+1/2) + \hbar g_m  \frac{(\hat{\sigma}_z+1)}{2}(\hat{b}+\hat{b}^\dagger),
\end{equation}
with $\hat{\sigma}_z=\vert e \rangle \langle e \vert - \vert g \rangle \langle g \vert$, and $\hat{b}$  the phonon annihilation operator. The last term describes the strain-mediated coupling, characterized
by the  coupling rate $g_m$. It can be rewritten as $\hbar g_m  \frac{(\hat{\sigma}_z+1)}{2}\frac{\hat{x}}{x_{\mbox{\scriptsize{zpf}}}} $, with $x_{\mbox{\scriptsize{zpf}}} =\sqrt{\hbar/2m\Omega_m}$  the zero point fluctuations amplitude of an oscillator of effective mass $m$.

\begin{figure}
\includegraphics[width=0.25\textwidth]{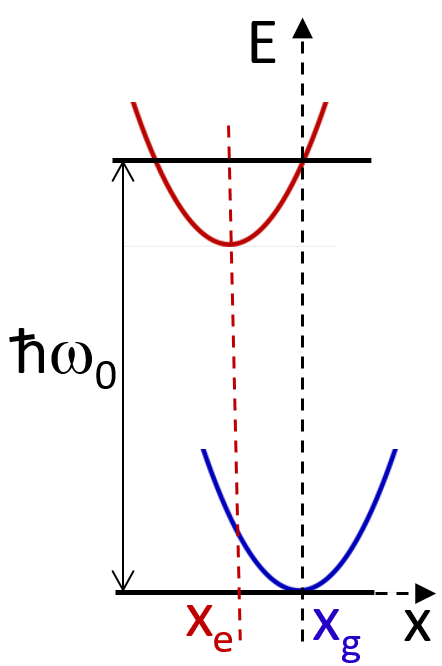}
\caption{\textbf{QD two-level scheme}. When the QD is in its ground state, its mechanical rest position is x$_g$. When the QD is excited, the linear strain-mediated potential shifts the oscillator rest position to x$_e$. }
\label{Harmonics}
\end{figure}

When the QD is in its ground state, the energy of the (QD+wire) system is given by the harmonic potential
\begin{equation}
 E_g=\frac{1}{2}m\Omega_m^2 (\mbox{x}-\mbox{x}_g)^2,
 \end{equation}
where x$_g$ is the rest position of the wire when the QD is in its ground state.   It is represented by the blue trace in Fig.~\ref{Harmonics}.
 
 When the QD is excited, the linear strain-mediated coupling term  is added to the harmonic potential, resulting in a displaced harmonic potential 
\begin{eqnarray}
 E_e &=& \hbar\omega_0+ \frac{1}{2}m\Omega_m^2 (\mbox{x}-\mbox{x}_g)^2 + \hbar g_m  \frac{(\langle \hat{\sigma}_z \rangle +1)}{2}\frac{\mbox{x}}{x_{\mbox{\scriptsize{zpf}}}} \\
 &=& \hbar \left(\omega_0 -\frac{g^2}{\Omega_m}\right) + \frac{1}{2}m\Omega_m^2 (\mbox{x}-\mbox{x}_e)^2
 \end{eqnarray} 
 with its minimum x$_e$ shifted (red trace in Fig.~\ref{Harmonics}) by 
\begin{equation}
\mbox{x}_e-\mbox{x}_g = -2\frac{g_m}{\Omega_m}   x_{\mbox{\scriptsize{zpf}}}.
\end{equation}


When the QD is driven by a saturating resonant laser on-off modulated  at the mechanical frequency, the mechanical oscillator sees its rest position  periodically displaced from x$_g$ to x$_e$, resulting in a periodic force
\begin{equation}
F_{\mathrm{QD}}(t) = \frac{1}{2} k(\mbox{x}_e-\mbox{x}_g) H(t) ,
\end{equation}
where the $1/2$ factor accounts for the average excited state QD population  at saturation $ \langle \hat{\sigma}_z \rangle = 0 $, $k=m\Omega_m^2$ is the oscillator stiffness, and $H(t)$ is the on-off periodic function equal to $1$ for half a period, and to $0$ for the next half-period. The Fourier decomposition of this function is
\begin{equation}
H(t) = \frac{1}{2} +\frac{2}{\pi} \sin \Omega_m t + \cdots  . 
\end{equation}

Note that this force can also be derived directly from the Hamiltonian as
\begin{equation}
F_{\mathrm{QD}}(t) = \frac{\partial \langle H \rangle}{\partial \mbox{x}}=\hbar g_m \frac{\langle \hat{\sigma}_z \rangle (t)+1}{2 x_{\mbox{\scriptsize{zpf}}}} .
\end{equation}
A periodic driving of the QD leads to a periodic QD population $\langle \hat{\sigma}_z\rangle (t)$ and therefore to a periodic force.

\subsection{Amplitude of the motion}
So the  amplitude of the QD induced  motion at resonance frequency $\Omega _m/2\pi$ is
\begin{eqnarray}
\mbox{x}_{\mathrm{QD}}^a&=& \frac{k(\mbox{x}_e-\mbox{x}_g)}{\pi m \Omega_m  \Gamma_m}
\\
&=&\frac{2}{\pi}\frac{g_m}{\Omega_m} Q x_{\mbox{\scriptsize{zpf}}},
\label{xQH}
\end{eqnarray}
where $\Gamma_m$ is the mechanical damping rate, and $Q=\frac{\Omega_m}{\Gamma_m}$ the mechanical quality factor. Note that the quantity experimentally measured and presented in the main text is the root mean square (rms) value and is given by  
\begin{equation}
x_{\mathrm{QD}}=\frac{\mbox{x}_{\mathrm{QD}}^a}{\sqrt{2}}=\frac{\sqrt{2}}{\pi}\frac{g_m}{\Omega_m} Q x_{\mbox{\scriptsize{zpf}}}.
\label{xQHrms}
\end{equation}

\subsection{Power}
Let us now calculate the average power of this force to recover the expression given in reference \cite{Auffeves}.
The power is  non-zero  only during the half period when the constant QD induced force $F_{\mathrm{QD}}^{cst}=k(\mbox{x}_e-\mbox{x}_g)/2$ applies. During this half period, the  wire top facet moves by $2x_{\mathrm{QD}}^a$ so that its average speed is $v=(2/\pi)\Omega_m \mbox{x}_{\mathrm{QD}}^a$.
The average power during a full period is then given by
\begin{equation}
P_{\mathrm{QD}}= \frac{1}{2}F_{\mathrm{QD}}^{cst} v = \frac{\hbar}{\pi^2}Qg_m^2 ,
\label{PQH}
\end{equation}
which is in line with reference \cite{Auffeves}.

\subsection{Alternative derivation}

This subsection presents an alternative derivation of the motion amplitude  and power based on energy considerations. 

In the stationary state, during the half-period during which the laser is on ($\langle \hat{\sigma}_z \rangle=0$), the  QD releases an energy $\Delta E_{\mathrm{QD}}=\hbar g_m\frac{\mbox{x}_{\mathrm{QD}}^a}{x_{\mbox{\scriptsize{zpf}}}}$ to the mechanical oscillator, and this energy is dissipated by the mechanical oscillator losses that are given  by $\Delta E_{\mathrm{dis}}= 2 \pi E_{\mathrm{tot}}/Q$ over a period, with $E_{\mathrm{tot}}= \frac{\hbar\Omega_m}{4} \left(\frac{\mbox{x}_{\mathrm{QD}}^a}{x_{\mbox{\scriptsize{zpf}}}}\right)^2$ the total energy of the system. In the stationnary state, we can  write
\begin{eqnarray}
\Delta E_{\mathrm{QD}}&=& \Delta E_{\mathrm{dis}}, \\
\hbar g_m\frac{\mbox{x}_{\mathrm{QD}}^a}{x_{\mbox{\scriptsize{zpf}}}} &=& \pi \frac{\hbar\Omega_m}{2} \left(\frac{\mbox{x}_{\mathrm{QD}}^a}{x_{\mbox{\scriptsize{zpf}}}}\right)^2  \frac{1}{Q}  
\end{eqnarray}
which leads to 
\begin{equation}
\mbox{x}_{\mathrm{QD}}^a= \frac{2}{\pi}\frac{g_m}{\Omega_m} Q x_{\mbox{\scriptsize{zpf}}}
\end{equation}
just as equation (\ref{xQH}) obtained above.

The average power of the QD induced motional drive is also simply given by
\begin{eqnarray}
P_{\mathrm{QD}} &=& \Delta E_{\mathrm{QD}}\frac{\Omega_m}{2\pi} \\
&=& \frac{\hbar}{\pi ^2} Q g_m^2 ,  
\end{eqnarray}
just as equation (\ref{PQH}) above.

\section{Wire geometry}

\begin{figure}
\includegraphics[width=0.3\textwidth]{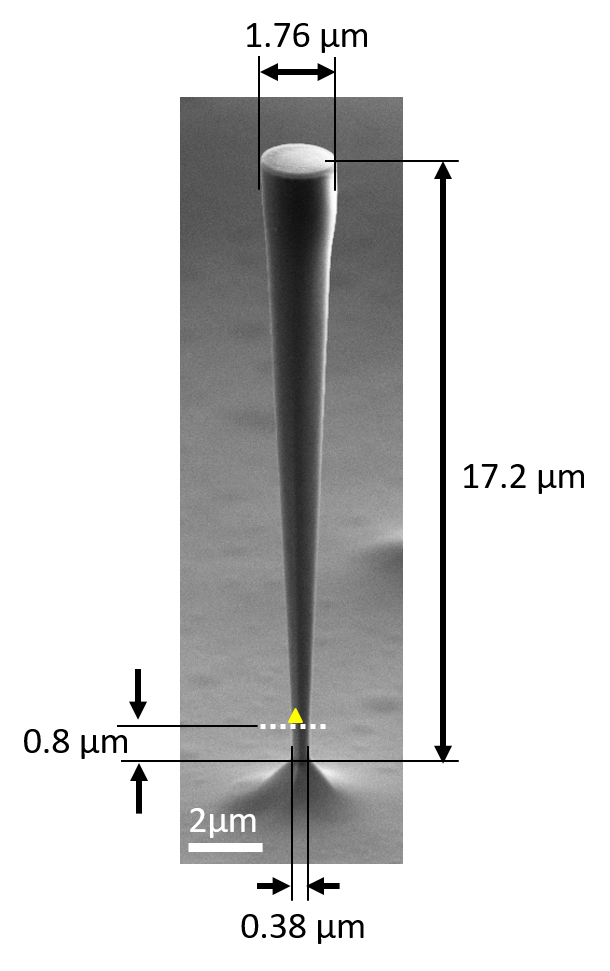}
\caption{\textbf{Wire dimensions}. The wire is made of GaAs. The InAs quantum dot is represented by a yellow triangle. }
\label{Tdim}
\end{figure}

The wire dimensions are displayed in Fig.~\ref{Tdim}: top facet radius $r_{\mathrm{top}}=0.88 \;\mu$m,
base radius $r_{\mathrm{base}}=0.19\;\mu$m, and height $h=17.2\;\mu$m. The wire volume is given by
\begin{equation}
V=\frac{\pi}{3}\frac{h}{(r_{\mathrm{top}}-r_{\mathrm{base}})}(r_{\mathrm{top}}^3-r_{\mathrm{base}}^3), 
\end{equation}
which gives $V=17.6\;\mu m ^3$. Knowing the volumic mass of GaAs $\rho=5.3$ g/cm$^3$, we obtain a geometric mass of $m_{\mathrm{geo}}=93$ pg. Note that, from Browian motion measurements, we have measured a effective mass $m=32 \;\mbox{pg}\;=\; 0.34 \; m_{\mathrm{geo}}$ (see section \ref{Sample_temperature}). As expected, this ratio is larger than  for the free-end motion of a singly clamped cylinder, for which the effective mass is $m=0.24 \; m_{\mathrm{geo}}$.

\section{Phase shift measurement}
\label{phase_shift}

\begin{figure*}
\includegraphics[width=0.75\textwidth]{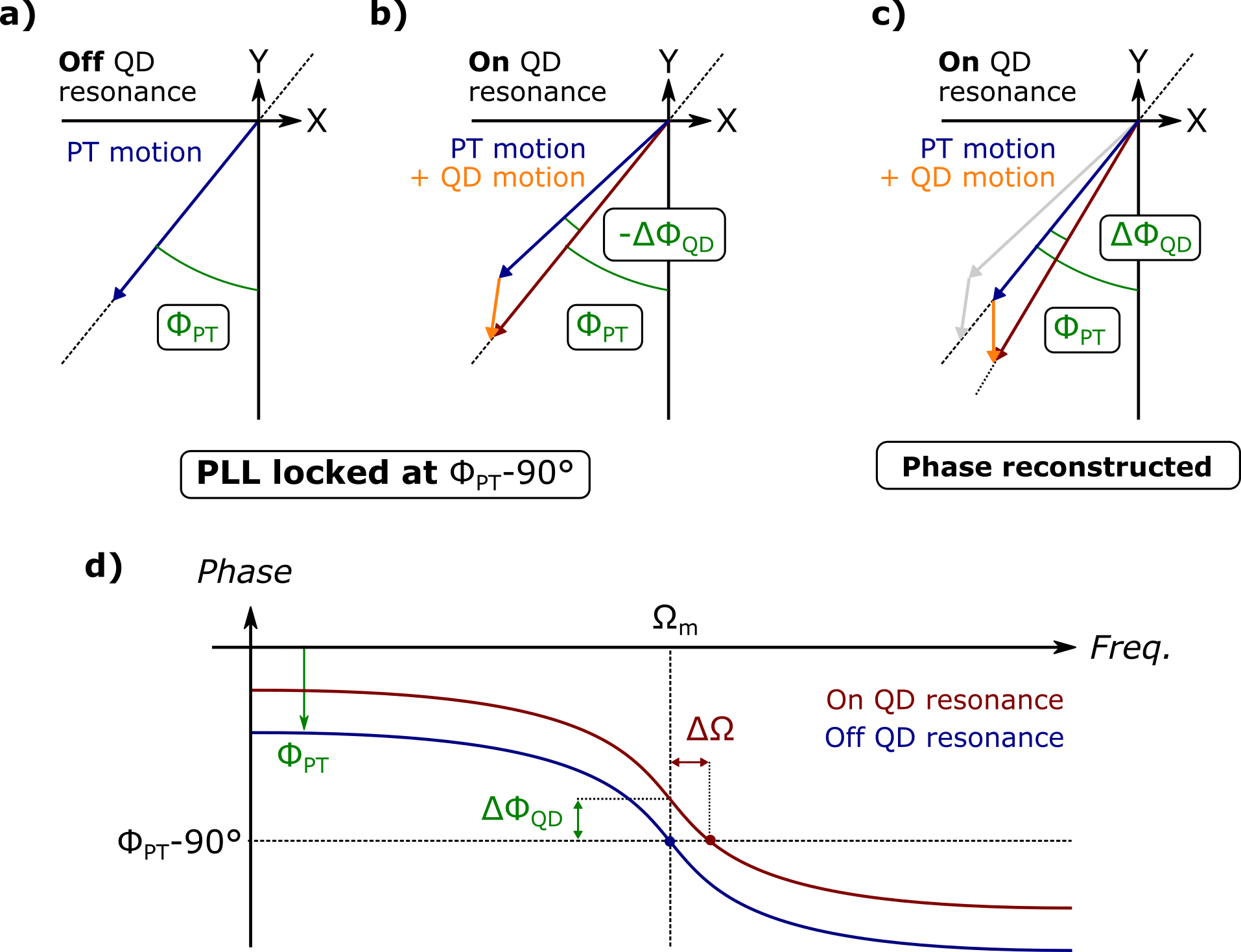}
\caption{\textbf{Phase measurement.} Representation of the motion vectors at mechanical resonance in the quadrature plane. 
 This sketch represents the coherent component of the wire motion  in the quadratures plane (rotating frame). $X$ and $Y$ are the rms values of the coherent  motion quadratures, 
defined as $x_c(t)=\sqrt{2}( X \cos \Omega_m t - Y \sin \Omega _m t )$. The phase reference is the laser modulation $I_{\mbox{las}}(t)= (I_0/2)(1+ \mbox{sgn}( \cos \Omega_m t))$ where $I_0$ is the laser intensity, and sgn the sign function (sgn$(x)=1$ ($-1$) if $x>0$ ($x<0$)). A phase delay is thus described with  a negative angle. 
  At mechanical resonance with a force responding instantaneous to the laser drive, the motion vector would be along the Y axis pointing towards $-\infty$.  \textbf{a}, When the laser is off resonance with the QD, only the photothermal (PT) motion is present and shown in blue at an angle $\Phi_{\mathrm{PT}}=-36^\circ$. The PLL is set to this angle to maintain the mechanical drive on resonance.   \textbf{b}, When the laser is on resonance with the QD, the QD induced motion (orange vector) adds to the PT motion. The  total motion is displayed as the dark red vector. The phase of the total motion is maintained by the PLL to $\Phi_{\mathrm{PT}}-90^\circ$ by changing the drive frequency (see text and \textbf{d}). In \textbf{c}, we show the total motion phase, reconstructed from the information of the PLL frequency change. It brings  the PT motion back along the $\Phi_{\mathrm{PT}}$ angle and the QD induced motion appears as changing the phase of the total motion by $\Delta \Phi_{\mathrm{QD}}$. This is what would happen with the PLL switched off, ie without the PLL locking the total phase to $\Phi_{\mathrm{PT}}$. \textbf{d}, Phase versus frequency plot across the mechanical resonance $\Omega _m $, exhibiting the $180^\circ$ phase difference between a frequency drive below and above resonance, with a $-90 ^\circ$ phase shift at resonance. The blue trace represents the photothermal  phase  (laser off QD resonance). It features a phase shift  $\Phi_{\mathrm{PT}}$ with respect to an instantaneous motional response. The operating point is the blue point of coordinates($\Omega_m, \Phi_{\mathrm{PT}}-90^\circ$). The dark red trace corresponds to the total motion when the QD induced motion is present (laser on QD resonance). It is phase shifted by $\Delta \Phi_{\mathrm{QD}}$ with respect to the PT motion alone. In this case, the PLL maintains the system to the set phase $\Phi_{\mathrm{PT}}-90^\circ$ by shifting the driving frequency by $\Delta \Omega$. The operating point becomes the dark red point of coordinates ($\Omega_m+\Delta \Omega, \Phi_{\mathrm{PT}}-90^\circ$).}
\label{phase}
\end{figure*}

We explain in this section how we measure the phase shift on the total  coherent motion caused by the QD induced motion. The total coherent wire motion is the sum of the photothermal (PT) motion (amplitude $\sim 50$ pm, phase shift $\Phi_{\mathrm{PT}}=-36^o$) and the QD induced motion (amplitude $\sim 0.6$ pm, quasi-instantaneous), as shown in Fig.~\ref{phase}. The photothermal effect does not depend on the laser detuning, whereas the QD induced motion appears only when the laser is resonant with the QD. The resonance of the laser with the QD has two effects: it increases the motion amplitude, and it shifts the phase of the total mechanical motion by $\Delta\Phi_{\mathrm{QD}}$.
To measure this phase shift 
we use the fact that the total mechanical motion phase is locked  via a phase locked loop (PLL) to a given set phase. In our case the PLL set phase is $\Phi_{\mathrm{PT}}-90^\circ$ corresponding to the phase  at mechanical resonance of the photothermal motion alone (see Fig.~\ref{phase}a,b,d). In usual operation, the PLL deals with slow mechanical frequency drifts by adjusting the mechanical oscillator driving frequency to track the set phase and therefore remain at the mechanical resonance. In our case, the PLL brings also an extra very useful information. During a laser scan across the QD resonance, when the laser becomes resonant with the QD optical transition,   the QD induced motion sets in and wants to modify the total motion phase by $\Delta\Phi_{\mathrm{QD}}$. But the PLL maintains the phase of the total motion to the set phase (see Fig.~\ref{phase}b) by correcting the driving frequency by $\Delta \Omega /2\pi=\frac{d\Omega /2\pi}{d\Phi}\Delta\Phi_{\mathrm{QD}}$, with $ d \Omega / d\phi =\Omega_m /(2Q)$ in the limit of $\Delta \Omega \ll \Omega_m/(2Q)$. When the QD induced motion is "on" at the QD resonance, the operating point is the dark red point  corresponding to coordinate  $\Omega _ m +\Delta\Omega$ and  $\Phi_{\mathrm{QD}}-90^\circ $ in Fig.\ref{phase}d. By recording the frequency change  $\Delta \Omega /2\pi$ of the PLL along a laser scan, we can infer the phase shift $\Delta\Phi_{\mathrm{QD}}$ caused by the QD induced motion as in Fig.~3c of the main text, and reconstruct the phase of the total motion as shown in Fig.~\ref{phase}c of the SI, and Fig.~3d of the main text.

\section{Photothermal effect}







 

\begin{table}
\begin{tabular}{|l|c|}
\hline
Heat capacity & $c=5\cdot 10^{-3} $ J/(g\cdot K)  \\
\hline
Thermal conductivity & $k=30$ W/(cm\cdot K)  \\
\hline
Volumic mass & $\rho=5.3$ g/cm$^3$ \\
\hline
Sound velocity & $v_s= 4000$ m/s \\
\hline
\end{tabular}
\caption{GaAs material data at $T=25$ K.}
\label{GaAs}
\end{table}

\subsection{Time scale of the effect}

When a thermal conductor has one of its dimensions much smaller than the mean free path of the bulk phonons, reduced size effects must be taken into account when estimating the thermal conduction and thus the characteristic thermalization time. This thermal conductivity must be renormalized by the value of the mean free path of the phonons in the nanostructure with respect to the bulk \cite{Bourgeois}. Relevant material data for GaAs are given in Table \ref{GaAs}.

Let's first calculate the bulk phonon mean free path $\Lambda_{ph}^{b}$ from the expression $k=(1/3)\rho c v_s \Lambda_{ph}^{b}$. We find $\Lambda_{ph}^{b}=0.1$ mm, which is much larger than the  wire dimensions. In the wire, we take, as a rough estimate, the mean free path as $\Lambda_{ph}^{w}=1\mu$m, so, from Casimir model, the effective thermal conductivity of the wire is $k^{w}=k \Lambda_{ph}^{w} / \Lambda_{ph}^{b}=0.3$ W/(cm.K).

The characteristic time for heat propagation in the wire is given by 
$\tau = \rho c V/ K_{w}$, with $V$ the wire volume and $K^{w}= k^{w} S/l$, the wire thermal conductance, with $S=0.25\mu$m$^2$ the wire section and $l=20\mu$m  its length. We find
 $\tau = 0.14 \;\mu$s. At $\Omega_m/2\pi=330$ kHz, the mechanical period is $T_m=3\;\mu$s. The characteristic time for heat propagation
  corresponds therefore to about $5\%$ of the mechanical period, ie to a delay of $\sim -20^o$. This value gives an order of magnitude which is not too far from the phase $\Phi_{\mathrm{PT}}=-36^o$ measured experimentally.
 





\subsection{Mitigation options}

A first solution would be to use frequency instead of intensity modulation to help mitigating the photothermal effect. We considered this solution early on in the project, but its technological implementation at such a modulation frequency ($330$ kHz) and over this optical frequency span ($5$ GHz), remains nowadays extremely challenging.

Another solution is to shift up the mechanical frequency to higher values. As is shown by our measurement, the photothermal effect is characterized by a thermalization time of $\Phi_{\mathrm{PT}}/ \Omega_m \sim0.3 \; \mu$s, such that if the mechanical frequency is larger, the photothermal effect will be averaged out and will vanish. This can be done e.g. by working with a smaller mechanical oscillator (with the advantage that one can in addition enhance $g_m$  significantly: see e.g.\cite{Artioli}), or by addressing a higher lying mechanical mode such as the vertical breathing mode ($\Omega_m / 2 \pi =29$ MHz) in our current oscillator.

Moreover, as photothermal effects are mainly due to crystalline and surface defects, it should certainly be possible to improve on this by developing specific bulk and surface treatments \cite{Guha}.

\section{Drifts and fluctuations}

Our experimental set-up  suffers from various drifts and fluctuations, that have to be dealt with as the experimental data is integrated over a time span of more than 30 hours.

\subsection{Mechanical resonance}

The wire is in vacuum sitting on the cold finger of the cryostat.  Owing to the imperfect vacuum in the cryostat, the wire  undergoes deposition of nitrogen and oxygen. This icing affects the mechanical properties of the wire. This leads to an increase of the wire's mass, but also stiffens it, so that the net effect on the mechanical frequency is altered but its evolution can not be anticipated. To deal with this drift, we use a Phase Locked Loop (PLL) to lock the driving frequency to the mechanichal resonance. This PLL is also used to extract the phase shift induced by the QD induced motion as explained in section \ref{phase_shift}.

Additionally, during the $30$ hours of data acquisition, this deposition  degrades the quality factor from $Q=2,700$ to $Q=1,500$, with an average of $Q=1,650$.
This Q factor drift occurs on a time scale slower ($ \frac{1}{Q} \frac{\delta Q }{\delta t}\sim 10^{-3}\;$ min$^{-1}$) than the laser sweep time (1 min), and Q can therefore be considered  constant during the duration of a laser sweep. 

Since the total coherent displacement is
\begin{equation}
\delta x[\Omega_m ]\propto \frac{Q}{m\Omega_m^2}(F_{\mathrm{PT}}+F_{\mathrm{QD}}),
\end{equation}
with $F_{\mathrm{PT}}$ and $F_{\mathrm{QD}}$ the photothermal and QD-induced force, respectively, the ratio between the QD-induced and the photothermal displacements is equal to the ratio between the QD-induced and the photothermal forces and is Q-independent. Our experimental procedure is to extract the Q-independent quantity $\delta x/Q$ for each sweep, and average these laser frequency dependent quantities over all the sweeps. The quantity plotted in Fig.3b of the main text is this average quantity multiplied by the average Q, that is $Q=1,650$.

\subsection{QD optical transition}
\label{wandering}

\begin{figure}
\includegraphics[width=0.47\textwidth]{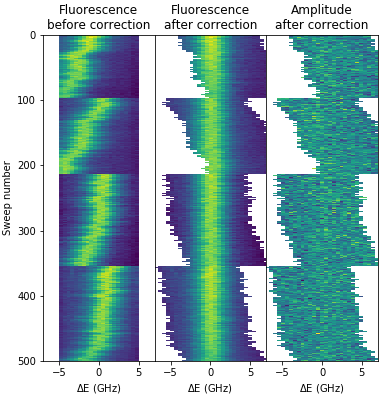}
\caption{\textbf{Spectral wandering compensation}. 
 On the left panel, each line is a 1 minute sweep of the resonant laser across the QD line. The three jumps around sweeps $\#100$, $\#210$ and $\#350$ correspond to moments where the QD line has drifted out of the laser sweep range, and a repositionning of the laser sweep range has been performed. For the left panel the zero of the $x$ axis is arbitrary. The central panel corresponds to the photoluminescence intensity after the recentering process in which each QD line has been fitted and aligned to a common center set to $\Delta E =0$.  On the right panel, the recentering procedure according to the photoluminescence data has been applied as well to the motion amplitude data. On this data, the QD-induced motion signal is not visible on a single sweep and is revealed only when averaging all the sweeps as shown in Fig.3b of the main text. For the sake of clarity, this figure represents only part of the total sweeps.   }
\label{SWC}
\end{figure}

The QD optical transition is subject to both drift and spectral wandering, as illustrated in Fig. \ref{SWC}. A cooling cycle lasts 4 days. During the first 2 days, the QD transition drift is too large to run the experiment in good conditions. After this time, the QD line becomes more stable but is still subject to spectral wandering as can be seen in Fig.~\ref{SWC}. 
 To benefit from a constructive addition of each sweep, the QD line of each sweep is fitted to a Gaussian and repositionned to a common center. The same operation is done for the motion data (see Fig.~\ref{SWC}). All the recentered sweeps are then added revealing the peak of the final data of Fig.3 of the main text. 

\subsection{Position drifts}

The sample position drifts by about $0.5 \;\mu$m/hour with respect to the lasers. To lock the lasers to the sample position, an extra laser ($\lambda=980$ nm) is reflected from the wire's top facet onto a four quadrant photodiode.  This information is used by two feedback loops via a Proportional, Integral, Derivative (PID) modules to react on the microscope objective x,y positions with a piezo driven stage. This allows us to run the experiment for more than 30 hours with all lasers well positionned with respect to the wire. 

\subsection{Resonant laser power correction}

\begin{figure}
\includegraphics[width=0.47\textwidth]{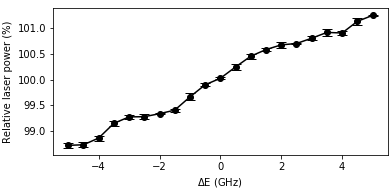}
\caption{
\textbf{Frequency dependent laser intensity}. 
Relative variation of the optical power during a laser sweep across the resonance of the QD. Average over 5 consecutive sweeps. The error bars are the standard error of the mean.}
\label{I_laser}
\end{figure}

The resonant laser power is changed by about $30\%$ as the laser frequency is swept across the QD line. This power change is partialy compensated using a feed-back loop acting of the Acousto-Optical Modulator (see experimental set-up in Fig.~2 of the main text). 
 After this correction, the resonant laser power still features a deterministic energy dependent intensity at the $1\%$ level as shown in Fig. \ref{I_laser}. As explained below (see section \ref{data}), this function is then used to renormalize the data to obtain a flat background.

\section{Phonon side band detection of the resonance fluorescence}

Detecting the QD resonance fluorescence directly on the QD line requires an efficient suppression of the back-reflected laser light using a cross polarization scheme.  But, owing to the degradation of the polarization of the laser light reflected from the wire top facet,  it is difficult to obtain a good polarisation rejection rate, which leads to reduced signal to background ratio.

\begin{figure}
\includegraphics[width=0.47\textwidth]{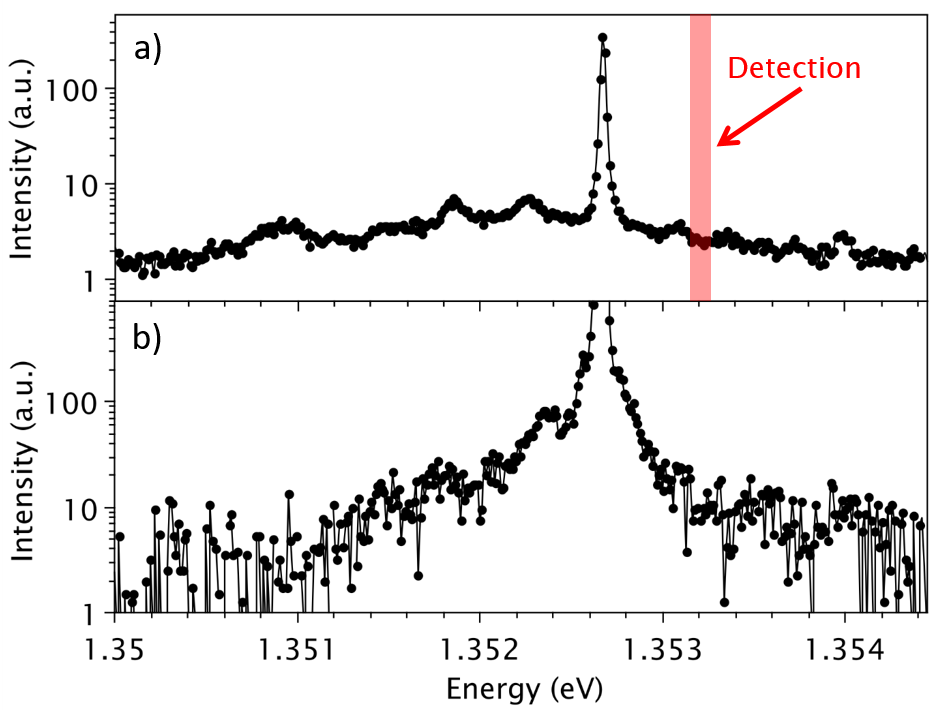}
\caption{\textbf{Detection on Phonon side band}. \textbf{a}, Photoluminescence spectrum, excited with the laser at $\lambda=825$ nm, showing the phonon side bands pedestal. For the QD induced motion experiment, the detection is performed at the energy represented by the pink zone over a bandwidth of $100\;\mu$eV.
\textbf{b}, Resonant laser spectrum, showing the presence of laser amplified spontaneous emission on the low energy side of the laser line.
}
\label{PSB}
\end{figure}

To avoid the back-reflected laser light, and improve further the signal to background ratio, we measure instead the fluorescence light on the high energy phonon side band  $0.5$ meV away from the QD line (see Fig.~\ref{PSB}a).
Let us recall that these high energy phonons ($0.5$ meV $=120$ GHz) are not at all related to the wire motion but to bulk semiconductor crystal thermal excitations.   
 The phonon side band signal is lower than the zero phonon line signal, but the laser background is almost completely removed leading to a greatly improved signal to noise ratio. We choose the apparently less favorable high energy phonon side-band, because of the presence of laser amplified spontaneous emission on the low energy side of the laser line (see Fig.~\ref{PSB} b).

\section{Data processing and error management}
\label{data}

\subsection{Data processing}

As explained in the Methods section of the paper, the data acquisition is performed by carrying out more than $1000$ one minute laser sweeps over the QD resonance, during an experimental time of more than $20$ hours. 

The total experimental run is split in $7$ blocks containing from $100$ to $1000$ sweeps. A new block is started whenever the QD resonance has drifted out of the laser scan range, and the laser scan range has been readjusted (see Fig. \ref{SWC} of the SI).

Before starting a new block $j$, the energy dependent laser intensity  is measured over $5$ sweeps and its average $I^j(\Delta E)$ (see Fig. \ref{I_laser}) is used to obtain a compensation function leading to a renormalized data with a flat background. The motion data $x_q^j(\Delta E)$ of  sweep $q$ of block $j$ is then corrected by this compensation function, while preserving the overall average, leading to the renormalized data data $^r x_q^j(\Delta E)$. The renormalized data is then recentered to correct for the spectral wandering (see Fig.  \ref{SWC}) as explained in section \ref{wandering} above. The quantity plotted in Fig.3b of the main text is the average of the renormalized and recentered data for all sweeps of all blocks.

\subsection{Error bars}

The error bars for each point in Fig. 3 of the main text represent the "standard error of the mean". They are calculated as the single shot signal variance divided by the square root of the total number of acquisitions. The error convergence behaviour has been experimentally verified by computing the root mean square error for various intermediate averaging durations $t_i$ expressed in number of binned sweeps units  (see Fig. \ref{SEM} below): A set of expected values 
$(\overline{X}_i)_k$
   of the observable  $X$ of interest (e.g. the amplitude, phase, frequency etc…) is computed over an acquisition time $t_i \ll T$, with $T$ the total acquisition time:
   
 \begin{equation}
 \overline{X}_{i,k}=\frac{1}{t_i} \int_{kt_i}^{(k+1)t_i} d\tau X(\tau ), \; k\in [0,\lfloor T/t_i \rfloor -1 ],
\end{equation}   
with $\lfloor T/t_i \rfloor $ the integer part of  $T/t_i$. 
The quantity $(\overline{X}_i)_k$
represents a random variable, with standard deviation 
$ (\sigma_X)_i$.
Fig. \ref{SEM}  below represents this standard deviation as a function of the averaging time $t_i$  (blue points). The red lines show a $1/\sqrt{t_i}$ law, which can be established following a calculation similar to that presented for the Brownian noise (see section \ref{Brownian} below). This shows that the data noise is dominated by Brownian noise and not by experimental stability and drift issues, and validates our error management. The discrepancy at large $t_i$ arises from the reduction of Card$[(\overline{X}_i)_k] $, which is why we extrapolate the error over the full acquisition length.

\begin{figure}
\includegraphics[width=0.47\textwidth]{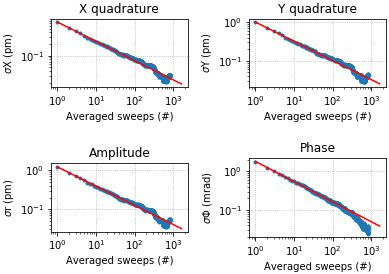}
\caption{\textbf{Error convergence}.
Error convergence as a function of the acquisition duration $t_i$ expressed in number of sweeps units. Each blue point is experimentally determined by calculating the standard deviation of the expected value acquired over the corresponding number of binned sweeps. The red line is an inverse square root law. 
}
\label{SEM}
\end{figure}

\section{Measurement of strain-mediated coupling strength $g_m$}

\begin{figure}
\includegraphics[width=0.4\textwidth]{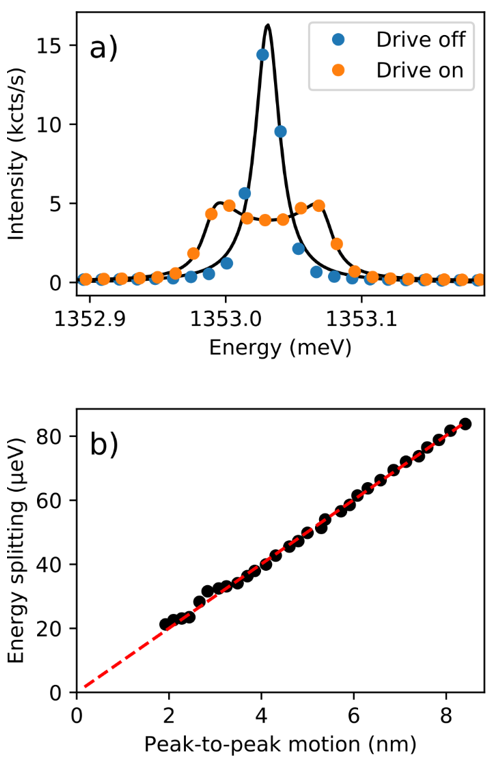}
\caption{\textbf{Measurement of strain coupling $g_m$}.
\textbf{a}, Photoluminescence spectrum of the QD line excited by the laser at $825$ nm, when the wire is not mechanically excited (blue dots, with a Lorentzian fit), and when it is mechanically excited by the photothermal effect of a modulated laser (orange dots). In the latter, the solid  line is a fit using a time averaged Lorentzian with a sinusoidally oscillating center. \textbf{b}, Mechanically induced splitting   as a function of motion amplitude. The slope is $\frac{dE}{dx}=\frac{\hbar g_m}{x_{\mbox{\scriptsize{zpf}}}}$ }
\label{g_measurement}
\end{figure}

The strain-mediated coupling strength $\hbar g_m=\frac{dE}{dx}x_{\mbox{\scriptsize{zpf}}}$ is obtained by measuring the spectral broadening $\Delta E$ of the QD photoluminescence spectrum as a function of the mechanical oscillation amplitude $\Delta x$ (see Fig.~\ref{g_measurement}). The oscillation amplitude is determined thanks to a careful motion calibration  as described in section \ref{calibration}. We measure $\frac{dE}{dx}=10.0 \,\mu$eV/nm (Fig.~\ref{g_measurement}b), so that $g_m/2\pi = 68 $ kHz.

\section{motion calibration}
\label{calibration}
Motion calibration relies on the careful determination of the coefficient $\beta=\frac{dV}{dx}$ giving the voltage change $\Delta V$ detected on the probe laser detector for a known displacement $\Delta x$. The calibrated displacement $\Delta x$ is performed by displacing the meter beam with a mirror controlled by a voltage $\Delta V_{\mathrm{PZT}}$ applied on a piezo-electrical transducer (PZT). The sample contains many wires arranged in a square lattice of parameter $a=15\;\mu$m, so, when applying a voltage $\Delta V_{\mathrm{PZT}}$ we can precisely measure the displacement $\Delta x$ of the beam on the sample by comparing it to the distance between two wires. This enables us to obtain $K=\frac{d x}{d V_{\mathrm{PZT}}}$. So by measuring $C=\frac{dV}{dV_{\mathrm{PZT}}}$, when the laser is close to the operating point (i.e. on the edge of the top facet of the wire with half of the power reflected) we can infer $\beta=C/K $. This measurement is carried out regularly, that is every two hours, during an experimental run to check that the value has not changed.

\section{Sample temperature}
\label{Sample_temperature}

\begin{figure}
\includegraphics[width=0.45\textwidth]{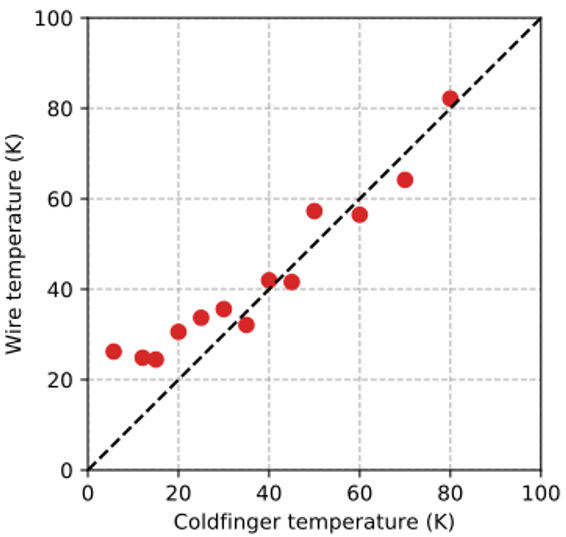}
\caption{\textbf{Brownian temperature versus cryostat temperature}. The wire temperature extracted from a Brownian motion measurement is plotted as a function of the temperature of the cold finger indicated on the cryostat. }
\label{TvsTcryo}
\end{figure}

We have checked the relation between the wire temperature extracted from the Brownian motion measurement and the cryostat temperature (Fig.~\ref{TvsTcryo}). The Brownian motion temperature, $T_{\mathrm{Br}}=\frac{\hbar \Omega_m}{2k_B}\frac{x_{\mathrm{Br}}^2}{x_{\mbox{\scriptsize{zpf}}}^2}$ with $k_B$ the Boltzmann constant, corresponds to the actual temperature of the wire $T_{\mathrm{wire}}=T_{\mathrm{Br}}$.
Above the cryostat temperature of $T_{\mathrm{cryo}}=40$K, the Brownian motion temperature is proportionnal to the cryostat temperature. The equality between these two temperatures is obtained when taking an effective mass $m=0.34 \, m_{\mathrm{geo}}=32$ pg.
As can be seen in Fig.~\ref{TvsTcryo}, the Brownian temperature is higher than the cryostat temperature at low temperature. At the operating cryostat temperature of $T_{\mathrm{cryo}}=5$ K, the actual temperature of the wire is $T_{\mathrm{wire}}=25$ K. This corresponds to the actual experimental conditions.

\section{Brownian motion averaging in coherent signal detection}
\label{Brownian}

In this section, we investigate the necessary time  to average out the incoherent Brownian motion in order to measure the coherent motion with a high enough signal to noise ratio to detect the rather weak QD induced motion.

Let us note $x_{\mathrm{sig}}(t)=X_{\mathrm{sig}}\cos{\Omega_mt}$ the coherent motion of frequency $\Omega_m/2\pi$ to be detected. We assume the measurement to be primarily limited by incoherent Brownian noise, $x_{\mathrm{Br}}(t)=X_{\mathrm{Br}}(t)\cos{\Omega_mt}+Y_{\mathrm{Br}}(t)\sin{\Omega_mt}$, with $X_{\mathrm{Br}}$ and $Y_{\mathrm{Br}}$ the quadratures of the Brownian motion. The total signal, proportional to $x(t)=x_{\mathrm{sig}}(t) +x_{\mathrm{Br}}(t)$, is  demodulated at $\Omega_m/2\pi$, the slowly varying component $X(t)$ of the resulting signal being retained:
\begin{eqnarray}
X(t)&=&X_{\mathrm{Br}}(t)+X_{\mathrm{sig}}\label{eq:1}
\end{eqnarray}
We assume that reducing the impact of the Brownian motion $X_{\mathrm{Br}}$ is performed by means of averaging. We subsequently define $s(\theta)$ as a linear, unbiased estimator for the measurement of $X_{\mathrm{sig}}$:
\begin{eqnarray}
s(\theta)=\frac{1}{\theta}\int_0^{\theta}\mathrm{d}t X(t), \label{eq:2}
\end{eqnarray}
with $\theta$ the measurement averaging time. $s(\theta)$ can be written as the sum of two contributions $s_{\mathrm{Br}}(\theta)$ and $s_{\mathrm{sig}}$ arising from the noise and signal, respectively:
\begin{eqnarray}
s_{\mathrm{Br}}(\theta)&=&\frac{1}{\theta}\int_0^{\mathrm{\theta}}\mathrm{d}\theta X_{\mathrm{Br}}(t),\nonumber\\
s_{\mathrm{sig}}&=&X_{\mathrm{sig}}.\label{eq:3}
\end{eqnarray}
The determination of a minimal value for $\theta$ requires defining a detection criterion. Here we will call the detection successful as soon as the dispersion of the estimator $\Delta s(\theta)=\sqrt{\Delta s^2(\theta)}$ is less than one fifth of the signal's expected value (also known as the 5-sigma criterion):
\begin{equation}
\Delta s(\theta) \leq \frac{s_{\mathrm{sig}}}{5}=\frac{X_{\mathrm{sig}}}{5}.\label{eq:4}
\end{equation}

The variance of $s$ arises from that of $s_{\mathrm{Br}}$, whose expected average value is zero-valued. One therefore has:
\begin{eqnarray}
\Delta s^2(\theta)&=&\langle s_{\mathrm{Br}}^2(\theta)\rangle\nonumber\\
&=&\frac{1}{\theta^2}\int_0^\theta\int_0^\theta\mathrm{d}t_1\mathrm{d}t_2\langle X_{\mathrm{Br}}(t_1)X_{\mathrm{Br}}(t_2)\rangle,\label{eq:5}
\end{eqnarray}
with $\langle...\rangle$ denoting statistical averaging. The integrand in Eq.(\ref{eq:5}) corresponds to the autocorrelation function of $X_{\mathrm{Br}}$. For Brownian motion, this autocorrelation function is stationary and only depends on the time difference $t_2-t_1$. Changing the variables $t_1$ and $t_2$ to $t'_1=t_1$ and $\tau=t_2-t_1$, one straightly obtains:
\begin{eqnarray}
\Delta s^2(\theta)&=&\frac{1}{\theta^2}\int_0^\theta\int_0^\theta\mathrm{d}t'_1\mathrm{d}\tau\langle X_{\mathrm{Br}}(0)X_{\mathrm{Br}}(\tau)\rangle\nonumber\\
&=&\frac{1}{\theta}\int_0^\theta\mathrm{d}\tau\langle X_{\mathrm{Br}}(0)X_{\mathrm{Br}}(\tau)\rangle.\label{eq:6}
\end{eqnarray}

The autocorrelation function of any given quadrature of the Brownian motion can be shown to have the following expression:
\begin{eqnarray}
\langle X_{\mathrm{Br}}(0)X_{\mathrm{Br}}(\tau)\rangle&=&\Delta x_{\mathrm{Br}}^2\times e^{-\Gamma_{\mathrm{m}}|\tau|/2},\label{eq:7}
\end{eqnarray}
with $\Gamma_{\mathrm{m}}$ the motion damping rate and $\Delta x_{\mathrm{Br}}^2$ the Brownian motion variance. Using this result, Eq.(\ref{eq:6}) yields to:
\begin{eqnarray}
\Delta s^2(\theta)&=&\frac{1}{\theta}\left[-\frac{2\Delta x_{\mathrm{Br}}^2}{\Gamma_{\mathrm{m}}}e^{-\Gamma_{\mathrm{m}}t/2}\right]_0^\theta\nonumber\\
&=&\frac{2\Delta x_{\mathrm{Br}}^2\left(1-e^{-\Gamma_{\mathrm{m}}\theta/2}\right)}{\Gamma_{\mathrm{m}}\theta}.\label{eq:8}
\end{eqnarray}
Assuming that averaging is required over a time much larger than the mechanical coherence time ($\theta\gg 1/\Gamma_{\mathrm{m}}$), the 5-sigma criterion (Eq.(\ref{eq:4})) yields to:
\begin{eqnarray}
\theta&\geq&\frac{50}{\Gamma_{\mathrm{m}}}\times\frac{1}{\mathrm{SNR}},\label{eq:9}
\end{eqnarray}
with $\mathrm{SNR}=\frac{X_{\mathrm{sig}}^2}{\Delta x_{\mathrm{Br}}^2}$ the signal-to-noise ratio. Eq.(\ref{eq:9}) shows that the minimum required averaging time is inversely proportional to the signal to noise ratio. In particular, it shows that for a SNR of $1$, successful detection is achieved after averaging for a minimal duration representing $50$ mechanical coherence times. In practice, other experimental aspects, such as technical noises and long term drifts, may represent even more critical limits in average-based measurement protocols.

\end{document}